# Variational self-consistent theory for trapped Bose gases at finite temperature

**Abd El-Aali Boudjemâa and Mohamed Benarous**

*Laboratory for Theoretical Physics and Material Physics*

*Faculty of Sciences and Engineering Sciences*

*Hassiba Benbouali University of Chlef*

*B.P. 151, 02000, Chlef, Algeria.*

**Abstract**

We apply the time-dependent variational principle of Balian-Vénéroni to a system of self-interacting trapped bosons at finite temperature. The method leads to a set of coupled non-linear time dependent equations for the condensate density, the thermal cloud and the anomalous density. We solve numerically these equations in the static case for a harmonic trap. We analyze the various densities as functions of the radial distance and the temperature. We find an overall good qualitative agreement with recent experiments as well as with the results of many theoretical groups. We also discuss the behavior of the anomalous density at low temperatures owing to its importance to account for many-body effects.

# Variational self-consistent theory for trapped Bose gases at finite temperature

## 1. Introduction

The Bose-Einstein condensation phenomenon was observed in dilute atomic gases for the first time in 1995[1, 2], and now becomes a very active area of research both theoretically and experimentally.

Weakly non-ideal Bose gases theory was pioneered by Bogoliubov [3]. The collective excitations turn out to be nicely predicted at and near zero temperature. Although there exist some generalizations to the non uniform case [4], the Bogoliubov approximation is primarily suited for homogeneous bose gases.

For trapped gases and at zero temperature, the Gross-Pitaevskii (GP) equation[5] constitutes the best model for studying the properties of Bose-Einstein condensation and provides excellent predictions for relevant experimental observations (see Dodd et al [6], Stringari [7], Proukakis et al. [8].) It is however only valid at $T$=0. Consequently, it ignores both the thermal cloud and the anomalous average and their dynamics.

At finite temperature, the mean-field approximation is intensively used to describe the static and dynamic equilibrium properties of Bose-Einstein condensates. Depending on the regime at hand, different variants and different models do exist. Among them, we can cite in particular the Hartree-Fock-Bogoliubov theory [9-11], the generalized GP equation and the Bogoliubov-de-Gennes (BdG) equations [12]. These theories have been successfully confronted to experiments as they predict correctly, among other things, the collective and single-particle excitations, the condensate fraction and also the transition temperature.

In order to take into account the motion of the condensate coupled to a static thermal cloud, Zaremba, Nikuni and Griffin [13] used the HFB-Popov approximation and the hydrodynamic approach in a kind of generalized mean field approximation. To go beyond, Proukakis[14], using an HFB basis, developed a second order perturbation theory to include the effect of the triplets.

Another kind of extensions has been developed by Stoof [15] and Gardiner and Zoller [16]. It is based on the Focker-Planck equation and the quantum optics techniques to derive a kinetic equation for the condensate coupled to a time dependent thermal cloud.

Yukalov[17] adopted a quite different formal approach by using the notion of representative statistical ensembles.

## Variational self-consistent theory for trapped Bose gases at finite temperature

The above theories, although being quite satisfactory, suffer from several drawbacks and inconsistencies which restrict their applicability to certain regimes and not to others. For instance, the Stoof, Gardiner-Zoller and Yukalov approaches ignore the dynamics of the anomalous density, despite its well established importance. On the other hand, the HFB approximation, which includes the anomalous average, leads to an unphysical gap in the excitation spectrum, which causes a violation of the Hugenholtz-Pines theorem [18, 19] and to UV divergences which require regularization. This has been done by Morgan [20] who implemented a ''generalized'' HFB approximation (GHFB) by resorting to the many-body T-matrix. It is not however not clear how to handle long range correlations which become more and more important as one approaches the critical regime. Moreover, going beyond the two-body contact potential is a cumbersome task in the GHFB approach. On the other hand, the perturbative approach is well suited in this respect but suffers from known shortcomings, especially near the critical regime.

In this paper we would like to proceed differently by using the time-dependent variational principle proposed by Balian and Vénéroni (BV) a long time ago [21]. The point is that this principle uses the notion of least biased state, which is the best ansatz compatible with the constraints imposed on the system. For our purposes, we will use a gaussian density operator. This ansatz is reminiscent of the independent particle approximation and belongs to the class of generalized coherent states. It allows us to perform explicitly all the calculations since there exists a generalization of Wick's theorem to statistical physics. It is important to note however that despite its simple form, it by no means endows any further simplification on the interaction between particles. Indeed, the BV-machinery, together with this trial gaussian ansatz, lead to a set of coupled time-dependent mean field equations for the condensate, the noncondensate and the anomalous average. We have to mention at this point that the equations that we derive in this paper are quite general and fully consistent as they do not require any simplifying assumption on the thermal cloud or the anomalous density. They may provide in this sense a kind of generalization to the previously discussed approximations, valid however, in the collisionless regime.

In their simplest (local) form, these equations have been derived elsewhere [22] and labeled TDHFB for Time Dependent Hartree-Fock-Bogoliubov. In the latter reference, these equations were solved numerically in the static case under the

# Variational self-consistent theory for trapped Bose gases at finite temperature

assumption that the spatial variations of the anomalous average were relatively small (which leads to a sort of finite temperature generalization of the Thomas-Fermi approximation). Although the results were quite satisfactory, at least on the qualitative level, there remained many unanswered questions. In particular, the thermal and anomalous density profiles seem to have no structure at the center of the trap. This is in contradistinction with what was found in the literature [11, 12] (where the HFB-BdG approximation was used), and where these densities are found to have ''holes''. On the other hand, a recent experiment [23] seems to indicate that the gaussian shape of the thermal cloud is maintained, therefore confirming the absence of structure near the center of the trap.

Hence, solving the full TDHFB equations, without any simplifying assumption, is of great interest in order to understand if the absence of these holes is an artifact of the finite temperature Thomas-Fermi approximation one adopted previously, or is a general prediction of the TDHFB approximation as a whole. In the latter case, one has to conclude definitely in favor of the predictions of the TDHFB approximation.

The paper is organized as follows. In section 2, we review the main steps used to derive the TDHFB equations from the Balian-Vénéroni variational principle. In section 3, the TDHFB equations are applied to a trapped Bose gas to derive a coupled dynamics of the condensate, the non condensate and the anomalous densities. We then restrict to the local densities and discuss the properties of the underlying equations, their relevance to the finite temperature case as well as their relations with other known approximations such as the Bogoliubov-De Gennes equations. In section 4, we present the static equations and the physical boundary conditions relevant to the trapped bose gas. Section 5 is devoted to the discussion of the numerical results, where we analyze the various density profiles. We confront our results with the HFB-BdG predictions and with recent experiments. Furthermore, owing to its importance to account for many-body effects, we focus on the behavior of the anomalous density for different temperatures. In section 6, we draw our conclusions and present some perspectives.

## 2. The variational TDHFB equations

# Variational self-consistent theory for trapped Bose gases at finite temperature

The time-dependent variational principle of Balian and Vénéroni requires first the choice of a trial density operator. In our case, we will consider a Gaussian time-dependent density operator. This ansatz which belongs to the class of the generalized coherent states allows us to perform the calculations since there exists a Wick's theorem, while retaining some fundamental aspects such as the pairing between atoms.

The Gaussian density operator $D(t)$ is completely characterized by the partition function $Z(t) = \mathrm{Tr}\, D(t)$, the one-boson field expectation values $<\Psi>(\vec{r},t) = \mathrm{Tr}\,\Psi(\vec{r})D(t)/Z(t)$, $<\Psi^+>(\vec{r},t) = \mathrm{Tr}\,\Psi^+(\vec{r})D(t)/Z(t)$ and the single-particle density matrix $\rho(\vec{r},\vec{r}',t)$ defined as

$$\rho(\vec{r},\vec{r}',t) = \begin{pmatrix} <\overline{\Psi}^+(\vec{r}')\overline{\Psi}(\vec{r})> & -<\overline{\Psi}(\vec{r}')\overline{\Psi}(\vec{r})> \\ <\overline{\Psi}^+(\vec{r}')\overline{\Psi}^+(\vec{r})> & -<\overline{\Psi}(\vec{r})\overline{\Psi}^+(\vec{r}')> \end{pmatrix}. \tag{2.1}$$

In the preceding definitions, $\Psi(\vec{r})$ and $\Psi^+(\vec{r})$ are the boson destruction and creation field operators (in the Schrödinger representation) satisfying the usual canonical commutation rules

$$\left[\Psi(\vec{r}),\Psi^+(\vec{r}')\right] = \delta(\vec{r}-\vec{r}'), \tag{2.2}$$

and $\overline{\Psi}(\vec{r}) = \Psi(\vec{r}) - <\Psi(\vec{r})>$ and $\overline{\Psi}^+(\vec{r}) = \Psi^+(\vec{r}) - <\Psi^+(\vec{r})>$ are the corresponding centered operators.

Upon introducing these variational parameters into the BV principle, one obtains

$$\begin{aligned} i\frac{\partial <\Psi>}{\partial t}(\vec{r},t) &= \frac{\partial E}{\partial <\Psi^+>(\vec{r},t)}, \\ i\frac{\partial <\Psi^+>}{\partial t}(\vec{r},t) &= -\frac{\partial E}{\partial <\Psi>(\vec{r},t)}, \\ i\frac{\partial \rho}{\partial t}(\vec{r},\vec{r}',t) &= -2\left[\rho,\frac{\partial E}{\partial \rho}\right](\vec{r},\vec{r}',t), \end{aligned} \tag{2.3}$$

where $E=<H>$ is the mean-field energy. These equations were also derived in [24]. Let us discuss some of their general properties. We note first that the mean-field energy $E$ is conserved when the Hamiltonian $H$ does not depend explicitly on time. Another property is the unitary evolution of the s.p. density matrix $\rho$, which means that the eigenvalues of $\rho$ are conserved. This implies in particular the conservation of the von-Neumann entropy $S = -\mathrm{Tr}\, D \log D$ and the fact that an initially pure state, satisfying $\rho(\rho+1)=0$, remains pure during the TDHFB evolution. The conservation of the free energy $F = E - TS$ naturally follows. These conservation laws show that the TDHFB equations reproduce, in the single-particle space, the exact properties of the system [24].

## 3. Application of the TDHFB formalism to trapped Bose gases

Let us apply the previous equations (2.3) to a system of trapped bosons interacting via a two-body potential. The grand canonical Hamiltonian may be written in the form

$$H = \int_{\vec{r}} \Psi^+(\vec{r}) \left[ -\frac{\hbar^2}{2M}\Delta + V_{\text{ext}}(\vec{r}) - \mu \right] \Psi(\vec{r}) + \frac{1}{2} \int_{\vec{r},\vec{r}'} \Psi^+(\vec{r}) \Psi^+(\vec{r}') V(\vec{r},\vec{r}') \Psi(\vec{r}') \Psi(\vec{r}'), \quad (3.1)$$

where $V(\vec{r},\vec{r}')$ is the interaction potential, $V_{\text{ext}}(\vec{r})$ the external confining field and $\mu$ the chemical potential. For the sake of clarity, we will omit to write explicitly the time dependence whenever evident. Next, we introduce the order parameter $\Phi(\vec{r}) = <\Psi(\vec{r})>$ and the non-local densities

$$\begin{aligned}\tilde{n}(\vec{r},\vec{r}') &\equiv \tilde{n}^*(\vec{r},\vec{r}') = <\Psi^+(\vec{r})\Psi(\vec{r}')> - \Phi^*(\vec{r})\Phi(\vec{r}'),\\ \tilde{m}(\vec{r},\vec{r}') &\equiv \tilde{m}(\vec{r}',\vec{r}) = <\Psi(\vec{r})\Psi(\vec{r}')> - \Phi(\vec{r})\Phi(\vec{r}').\end{aligned} \quad (3.2)$$

where we note that $\tilde{n}(\vec{r},\vec{r}) \equiv \tilde{n}(\vec{r})$ and $\tilde{m}(\vec{r},\vec{r}) \equiv \tilde{m}(\vec{r})$ are respectively the non condensate and the anomalous densities. The energy may be readily computed to yield

$$\begin{aligned}E = &\int_{\vec{r}} h^{sp}(\vec{r}) [\tilde{n}(\vec{r},\vec{r}) + \Phi(\vec{r})\Phi^*(\vec{r})] + \int_{\vec{r},\vec{r}'} V(\vec{r},\vec{r}') |\Phi(\vec{r})|^2 |\Phi(\vec{r}')|^2\\ &+ \frac{1}{2}\int_{\vec{r},\vec{r}'} V(\vec{r},\vec{r}') [\tilde{m}^*(\vec{r},\vec{r}')\tilde{m}(\vec{r},\vec{r}') + \tilde{n}(\vec{r},\vec{r}')\tilde{n}(\vec{r}',\vec{r}) + \tilde{n}(\vec{r},\vec{r})\tilde{n}(\vec{r}',\vec{r}')]\\ &+ \frac{1}{2}\int_{\vec{r},\vec{r}'} V(\vec{r},\vec{r}') [\tilde{n}(\vec{r},\vec{r}')\Phi(\vec{r})\Phi^*(\vec{r}') + \tilde{n}(\vec{r}',\vec{r})\Phi^*(\vec{r})\Phi(\vec{r}') + \tilde{n}(\vec{r},\vec{r})\Phi(\vec{r}')\Phi^*(\vec{r}') + \tilde{n}(\vec{r}',\vec{r}')\Phi(\vec{r})\Phi^*(\vec{r})]\\ &+ \frac{1}{2}\int_{\vec{r},\vec{r}'} V(\vec{r},\vec{r}') [\tilde{m}^*(\vec{r},\vec{r}')\Phi(\vec{r})\Phi(\vec{r}') + \tilde{m}(\vec{r},\vec{r}')\Phi^*(\vec{r})\Phi^*(\vec{r}')].\end{aligned}$$

(3.3)

In the equation (3.3), $h^{sp} = -\frac{\hbar^2}{2M}\Delta + V_{\text{ext}}(\vec{r}) - \mu$ is the single particle Hamiltonian.

Now, one inserts the expression (3.3) in the general equations of motion (2.3) to get the explicit form of the TDHFB equations for a trapped Bose gas:

# Variational self-consistent theory for trapped Bose gases at finite temperature

$$i\hbar\dot{\Phi}(\vec{r}) = h^{sp}(\vec{r})\Phi(\vec{r})$$
$$+ \int_{\vec{r}'} V(\vec{r},\vec{r}')\left[|\Phi(\vec{r}')|^2\Phi(\vec{r}) + \Phi^*(\vec{r}')\tilde{m}(\vec{r},\vec{r}') + \Phi(\vec{r}')\tilde{n}(\vec{r},\vec{r}') + \Phi(\vec{r})\tilde{n}(\vec{r}',\vec{r}')\right] \quad (3.4a)$$

$$i\hbar\dot{\tilde{n}}(\vec{r},\vec{r}') = \left[h^{sp}(\vec{r}) - h^{sp}(\vec{r}')\right]\tilde{n}(\vec{r},\vec{r}')$$
$$+ \int_{\vec{r}''} V(\vec{r}',\vec{r}'')\left[a(\vec{r}'',\vec{r}')\tilde{n}(\vec{r},\vec{r}'') + a(\vec{r}'',\vec{r}'')\tilde{n}(\vec{r},\vec{r}') + b(\vec{r}',\vec{r}'')\tilde{m}(\vec{r}'',\vec{r})\right] \quad (3.4b)$$
$$- \int_{\vec{r}''} V(\vec{r},\vec{r}'')\left[a(\vec{r},\vec{r}'')\tilde{n}(\vec{r}'',\vec{r}) + a(\vec{r}'',\vec{r}'')\tilde{n}(\vec{r},\vec{r}') + b(\vec{r},\vec{r}'')\tilde{m}(\vec{r}'',\vec{r}')\right],$$

$$i\hbar\dot{\tilde{m}}(\vec{r},\vec{r}') = \left[h^{sp}(\vec{r}) + h^{sp}(\vec{r}')\right]\tilde{m}(\vec{r},\vec{r}')$$
$$+ \int_{\vec{r}''} V(\vec{r}',\vec{r}'')\left[a(\vec{r}'',\vec{r}')\tilde{m}(\vec{r},\vec{r}'') + a(\vec{r}'',\vec{r}'')\tilde{m}(\vec{r},\vec{r}') + b(\vec{r}',\vec{r}'')\left(\tilde{n}^*(\vec{r},\vec{r}'') + \delta(\vec{r}-\vec{r}'')\right)\right]$$
$$+ \int_{\vec{r}''} V(\vec{r}',\vec{r}'')\left[a(\vec{r}'',\vec{r})\tilde{m}(\vec{r}',\vec{r}'') + a(\vec{r}'',\vec{r}'')\tilde{m}(\vec{r},\vec{r}') + b(\vec{r},\vec{r}'')\tilde{n}(\vec{r}'',\vec{r}')\right].$$
$$(3.4c)$$

In the Eqs. (3.4), the dots denote time derivatives and we have introduced the quantities

$$a(\vec{r},\vec{r}') = \tilde{n}(\vec{r},\vec{r}') + \Phi^*(\vec{r})\Phi(\vec{r}'),$$
$$b(\vec{r},\vec{r}') = \tilde{m}(\vec{r},\vec{r}') + \Phi(\vec{r})\Phi(\vec{r}').$$
$$(3.5)$$

It is worth noticing that similar equations have been derived elsewhere using quite different approaches. For instance, Stoof [25] used a variational plus perturbative effective action, Proukakis [14] a truncation of the Heisenberg equations and Chernyak et al. [26] the generalized coherent state representation. The latter approach yields equations very close to ours, but the authors did not pursue further their analysis. In particular, an important diverging term in their equations was completely ignored. In our case, this term appears in the equation (3.4c) and becomes highly non trivial when considers the contact potential. This is precisely the term which leads to UV- divergences in the anomalous density [12] and which requires a regularization. We do this by a very natural recipe which consists in defining the renormalized anomalous average by

$$\tilde{m}(\vec{r},\vec{r}') = \tilde{m}_R(\vec{r},\vec{r}') + \frac{1}{i\hbar}\int_t b(\vec{r},\vec{r}',t)\delta(\vec{r}-\vec{r}'), \quad (3.6)$$

and replacing $\tilde{m}(\vec{r},\vec{r}')$ by $\tilde{m}_R(\vec{r},\vec{r}')$ wherever it appears. This time-dependent renormalization turns out to be quite similar to the static regularization scheme introduced by Morgan [20] and also to the rigorous derivation of Olshanii and

# Variational self-consistent theory for trapped Bose gases at finite temperature

Pricoupenko[19, 27] based on the pseudo-potential method. In what follows, we will omit the index "R" and simply write $\tilde{m}$ instead of $\tilde{m}_R$.

The equations (3.4) remain complicated even for a contact potential. To proceed further, and by the same way, to investigate the behavior of the various density profiles, we will change to a local representation by taking the limit $\vec{r}' \to \vec{r}$. We further consider the contact potential $V(\vec{r},\vec{r}') = g\delta(\vec{r}-\vec{r}')$, where g is related to the s-wave scattering length $a$ by $g = 4\pi\hbar^2 a/M$. For an isotropic trap, and due to the spherical symmetry, the equations (3.4) become

$$i\hbar\dot{\Phi}(r) = \left(-\frac{\hbar^2}{2M}\Delta + V_{ext}(r) - \mu + gn_c(r) + 2g\tilde{n}(r)\right)\Phi(r) + g\,\tilde{m}(r)\Phi^*(r),$$

$$i\hbar\dot{\tilde{m}}(r) = g(2\tilde{n}(r)+1)\,\Phi^2(r) + 4\left(-\frac{\hbar^2}{2M}\Delta + V_{ext}(r) - \mu + 2gn(r) + \frac{g}{4}(2\tilde{n}(r)+1)\right)\tilde{m}(r),$$

$$i\hbar\dot{\tilde{n}}(r) = g\left(\tilde{m}^*(r)\Phi^2(r) - \tilde{m}(r)\Phi^{*2}(r)\right),$$

(3.7)

where $n_c(r) = |\Phi(r)|^2$ is the condensate density and $n(r)$ the total density $n(r) = n_c(r) + \tilde{n}(r)$.

It is important to note that the equations (3.7) are not totally independent. Indeed, $\tilde{n}$ and $\tilde{m}$ are related by the equality

$$(1+2\tilde{n})^2 - |2\tilde{m}|^2 = I,\qquad(3.8)$$

where $I$ is the Heisenberg parameter. At thermal equilibrium, it is related to the temperature by $\sqrt{I} = 1 + 2f_B(E)$, where $f_B$ is the Bose-Einstein distribution for a system of energy $E$. Hence, at $T=0$, $I=1$. As it stands, $I$ describes the degree of mixing of the system. It indeed obeys the continuity equation

$$\frac{\partial I}{\partial t} + \vec{\nabla}\cdot\vec{J}_I = 0,\qquad(3.9)$$

where the current density is $\vec{J}_I = -\frac{2i\hbar}{M}\left(\tilde{m}^*\vec{\nabla}\tilde{m} - \tilde{m}\vec{\nabla}\tilde{m}^*\right)$. The deviation from the pure state situation ($I=1$) is therefore mainly controlled by the anomalous density. This result shows that neglecting $\tilde{m}$ while maintaining $\tilde{n}$ will necessarily lead to inconsistencies when talking about the mixing of the condensed and non-condensed phases. Furthermore, the total density also obeys a continuity equation of the form

$$\frac{\partial n}{\partial t} + \vec{\nabla} \cdot \vec{J} = 0, \tag{3.10}$$

where $\vec{J} = -\frac{i\hbar}{2M}\left(\Phi^*\vec{\nabla}\Phi - \Phi\vec{\nabla}\Phi^*\right)$, which leads to the conservation of the total number of atoms $N = \int_{\vec{r}} n(r)$. It is also interesting to observe here that even at finite temperature, there is no current associated with the thermal cloud.

Let us now turn to discuss the relationship between our TDHFB equations (3.7) and the HFB-De Gennes equations [11-12]. In fact, we may easily show that upon linearizing the first equation in (3.7) around a static solution, we obtain

$$i\hbar\delta\dot{\Phi} = \left(h^{sp} + 2gn\right)\delta\Phi + g\left(\tilde{m} + \Phi^2\right)\delta\Phi^*. \tag{3.11}$$

Using the parametrization $\delta\Phi = \sum_p \left(u_p e^{-i\hbar\omega_p t} - v_p e^{i\hbar\omega_p t}\right)$ in which $\omega_p$ are the quasi-particle frequencies and $u_p$ and $v_p$ the quasi-particle amplitudes, we get from (3.11) the HFB-BdG equations:

$$\begin{aligned}\varepsilon_p u_p &= \left(h^{sp} + 2gn\right)u_p - g\left(n_c + \tilde{m}\right)v_p, \\ -\varepsilon_p v_p &= \left(h^{sp} + 2gn\right)v_p - g\left(n_c + \tilde{m}\right)u_p.\end{aligned} \tag{3.12}$$

Hence, the HFB-BdG equations found in the literature are just the random phase approximation to the first of our equations (3.7), which are therefore more general since the two last equations in (3.7) have no analogues in the HFB formalism.

## 4. The Static TDHFB Equations

The static TDHFB equations are obtained by setting to zero the right hand sides of (3.7). For numerical purposes, it is convenient to start our treatment with the dimensionless form of the set (3.7). Let us consider a spherical trap with frequency $\omega$, $V_{ext}(r) = \frac{1}{2}m\omega^2 r^2$ and use the harmonic oscillator length $a_{HO} = \sqrt{\hbar/m\omega}$, as well as $a_{H0}^{-3}$ and $\hbar\omega$ as units of length, density and energy respectively. The dimensionless radial distance is $q = r/a_{HO}$. The dimensionless condensed, non-condensed and anomalous densities are respectively $\hat{n}_c = a_{HO}^3 n_c$, $\hat{\tilde{n}} = a_{HO}^3 \tilde{n}$ and $\hat{\tilde{m}} = a_{HO}^3 \tilde{m}$. Therefore, $\hat{n} = \hat{n}_c + \hat{\tilde{n}}$ is the dimensionless total density. However, in order to avoid the appearance of first

## Variational self-consistent theory for trapped Bose gases at finite temperature

order spatial derivatives in the final equations (which are always cumbersome in a numerical treatment), it is preferable to introduce a dimensionless "order parameter" $\hat{\Psi} = q\hat{n}_c$ and a dimensionless "anomalous density" $\hat{\chi} = q\hat{\tilde{m}}$. With these definitions, the static equations corresponding to (3.7) write as:

$$\left(-\frac{d^2}{dq^2} + q^2 - \nu + 8\pi \frac{a}{a_{HO}}\left(\frac{\hat{\Psi}^2}{q^2} + 2\hat{\tilde{n}} + \frac{\hat{\chi}}{q}\right)\right)\hat{\Psi} = 0,$$

$$\left(-\frac{d^2}{dq^2} + q^2 - \nu + 8\pi \frac{a}{a_{HO}}\left(2\hat{n} + \frac{2\hat{\tilde{n}} + \hat{V}^{-1}}{4}\right)\right)\hat{\chi} + 8\pi \frac{a}{a_{HO}}\left(\frac{2\hat{\tilde{n}} + \hat{V}^{-1}}{4}\right)\frac{\hat{\Psi}^2}{q} = 0,$$

(4.1)

where $\hat{V} = V/a_{HO}^3$ is the normalized volume and we have introduced the dimensionless chemical potential $\nu = \mu / \frac{1}{2}\hbar\omega$. We have deliberately omitted the third equation in (3.7) since it is of no interest for our purposes. Indeed, the quantity $\hat{\tilde{n}}$ is determined via the relation (3.8) which becomes in our units

$$\hat{\tilde{n}} = \frac{1}{2}\left[\left(I + 4\frac{\hat{\chi}^2}{q^2}\right)^{1/2} - 1\right]. \qquad (4.2)$$

The atom-atom interaction is now completely specified by the parameter $a/a_{HO}$. The normalization condition writes in these new units as $\int d^3q\, \hat{n}(q) = N$, where $N$ is the total number of atoms.

As one notes from (4.1), the problem becomes formally one-dimensional. Its solution requires a set of boundary conditions as $r \to \infty$. We have therefore to determine the asymptotic behavior of $\Phi$ and $\tilde{m}$ or equivalently $\hat{\Psi}$ and $\hat{\chi}$. To this end, we note first that for a confined gas, $\Phi$ and $\tilde{m}$ must vanish at infinity. However, these boundary conditions are numerically useless since they always lead to the trivial solution. Hence, it is preferable to solve the equations (4.1) in the asymptotic region (where the non linear terms can be neglected) and then to consider these asymptotic solutions as boundary conditions for the problem (4.1). In the region $q \to \infty$, the two equations in (4.1) become identical and lead to the same generic solution : $\hat{\Psi} \approx \hat{\chi} \approx q\exp(-q^2)$.

Next we discretize the set of equations (4.1) by using a finite difference scheme with step $h$. The chemical potential must be computed self-consistently in order to ensure the normalization condition. One may associate with the algorithm a gradient method,

**Variational self-consistent theory for trapped Bose gases at finite temperature**

which searches for the best value of $\nu$ that satisfies the previous condition. However, this procedure is rather slowly converging and we prefer instead to use a more intuitive and less expensive technique. The idea is to start our calculations for $a/a_{H0} \ll 1$, in which case we know that the value of $\nu$ is very close to that of the ideal gas, namely $\nu = 3$ (or $\mu = \frac{3}{2}\hbar\omega$). By gradually increasing the ratio $a/a_{H0}$, we may correspondingly determine $\nu$.

## 5. Numerical Results

To illustrate our formalism at finite temperature, we consider the $^{87}Rb$ gas with $a/a_{H0} = 7.64\,10^{-3}$ and $N = 2000$ atoms.

We begin by plotting the condensate, the non condensate and the anomalous densities as functions of the radial distance for several values of the condensate fraction $N_c/N$ and hence of the temperature. Densities and lengths are measured in units of $a_{HO}^{-3}$ and $a_{HO}$ respectively.

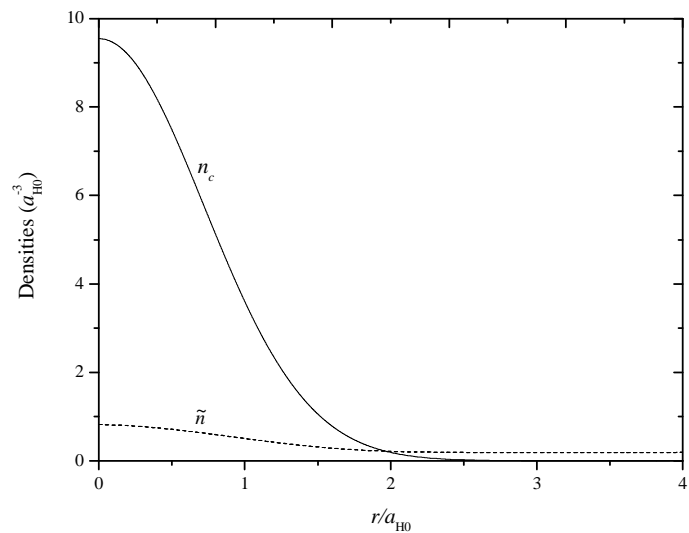

**Figure 1. The condensate (solid) and non-condensate (dashed) densities versus the radial distance (in units of $a_{H0}$) for: $a/a_{H0} = 7.64\,10^{-3}$, $N = 2000$ atoms and $N_c/N = 85\%$.**

In figure 1, we show the two components of the gas at low temperature. We notice that the non condensate density is rather small compared to the condensate density, since in this range of temperature, almost all the atoms are condensed in the center of

**Variational self-consistent theory for trapped Bose gases at finite temperature**

the trap. We also observe that the thermal cloud has a larger tail compared to the condensate. This image illustrates clearly the fact that the condensate is surrounded by the thermal cloud. The previous results have also been obtained by several authors, see e.g. [11, 12, 28].

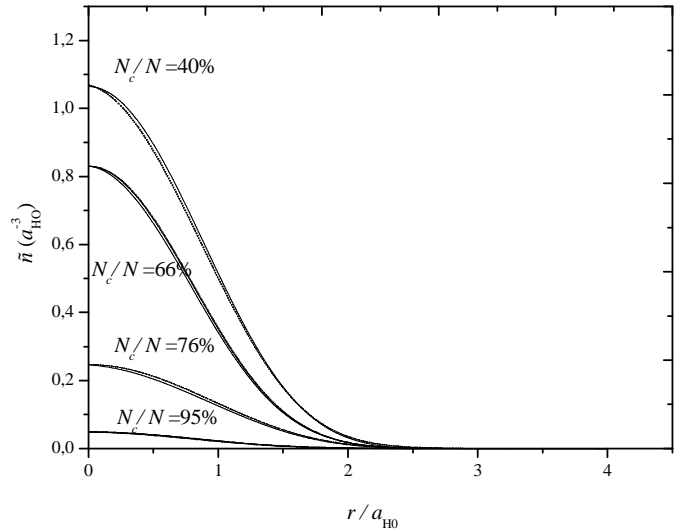

**Figure 2. The non-condensate density versus the radial distance for various condensate fractions, $a/a_{H0} = 7.64\,10^{-3}$ and $N = 2000$ atoms. Solid: our result, dashed: the ideal gas result.**

In the figure 2, we see that the non-condensate density is increasing significantly with decreasing condensate fraction. Furthermore, it is quite interesting to observe that the ideal gas model (dashed line) is still a good approximation.

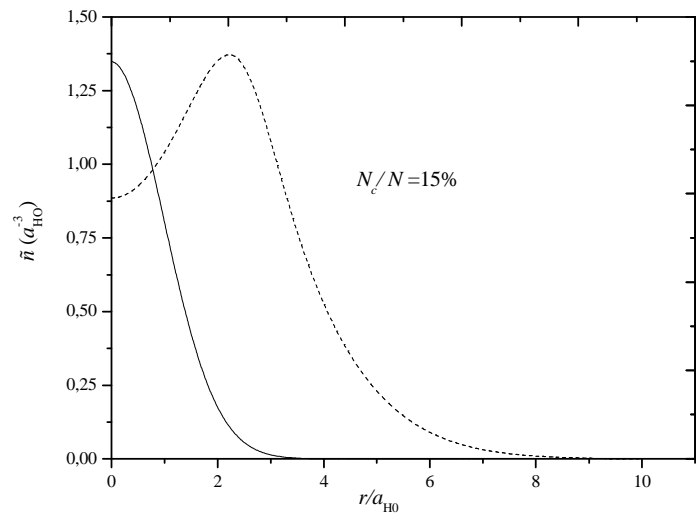

**Figure 3.** The non-condensate density versus the radial distance for $N_c/N = 15\%$, $a/a_{H0} = 7.64\,10^{-3}$ and $N = 2000$ atoms. Solid: our result, dashed: HFB-BdG.

In the figure 3, we compare our non condensate density with the one predicted by the HFB-BdG equations. What is important to note here is that we clearly disagree with the HFB-BdG calculations [12]. The discrepancies are more pronounced near the center of the trap, where the HFB-BdG approximation predicts a hole. On the other hand, our result is in accordance with the experiments of Ref. [23, 29], where no special structure is observed near the center and where the gaussian shape of the thermal cloud seems to be maintained. These observations may be extended to the anomalous average. Indeed, we see in the figure 4 that our results are again in clear disagreement with the HFB-BdG approximation which also predicts a hole in the center while we predict a gaussian shape.

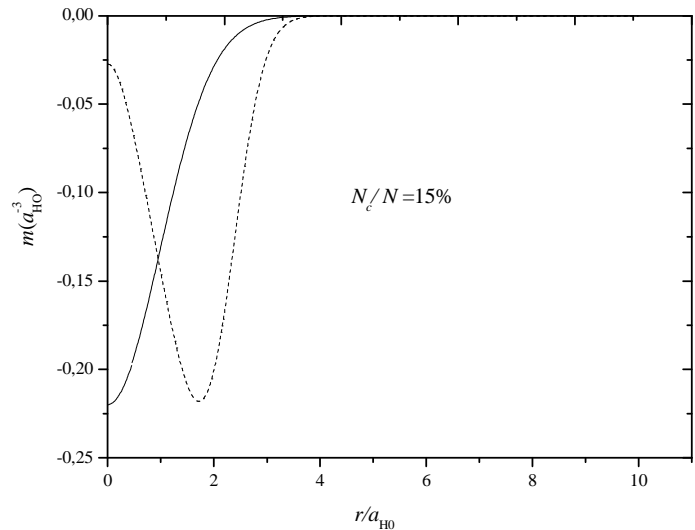

**Figure 4.** The anomalous density versus the radial distance for $N_c/N = 15\%$, $a/a_{H0} = 7.64\,10^{-3}$ and $N = 2000$ atoms. Solid: our result, dashed: HFB-BdG.

The figure 5 depicts the anomalous density for varying condensation fraction. We notice that by decreasing $N_c/N$, $\tilde{m}$ begins to increase in absolute value then decreases when $N_c/N$ approaches 50%. This overall behavior has also been obtained in [11, 12]. Therefore, our calculations predict no special structure of $\tilde{n}$ and $\tilde{m}$ near the center and a monotonic behavior from the center to the edges of the trap. In fact,

**Variational self-consistent theory for trapped Bose gases at finite temperature**

$\tilde{m}$ is still ill known quantity and an experiment (although highly non trivial) focusing on this quantity would be welcome in the future.

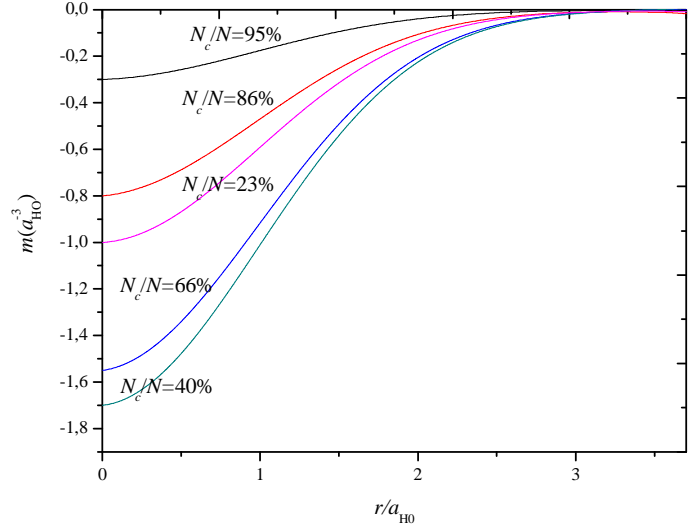

**Figure 5. The anomalous density versus the radial distance for various condensate fractions, $a/a_{H0} = 7.64\,10^{-3}$ and $N = 2000$ atoms.**

The figures 6, 7 and 8 compare the absolute value of the anomalous density $|\tilde{m}|$ (in black) and the non condensate density $\tilde{n}$ (in red). We observe that at high condensate fraction (that is at very low temperatures), for instance at $N_c/N = 95\%$, $|\tilde{m}|$ is greater than $\tilde{n}$, although both are very small. For $N_c/N \approx 50\%$ (Fig. 7), the anomalous density is comparable to the thermal cloud and at low enough condensate fraction (Fig. 8), the anomalous density becomes much smaller than the non condensate density. Hence, we may infer from that that the anomalous density plays a central role at low temperatures. It is therefore highly unlikely to neglect it for $T << T_c$. These observations are consistent with the results of Yukalov[17, 30].

**Variational self-consistent theory for trapped Bose gases at finite temperature**

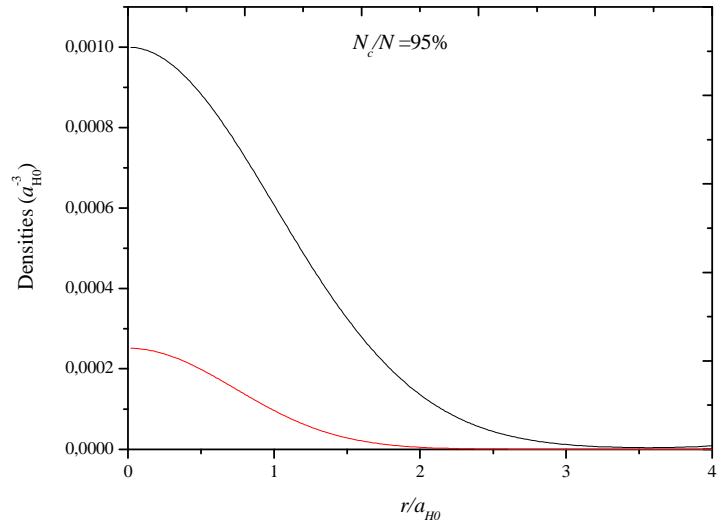

**Figure 6. The absolute value of the anomalous density (black) and the non condensate density (red) vs. the radial distance for $N_c/N = 95\%$.**

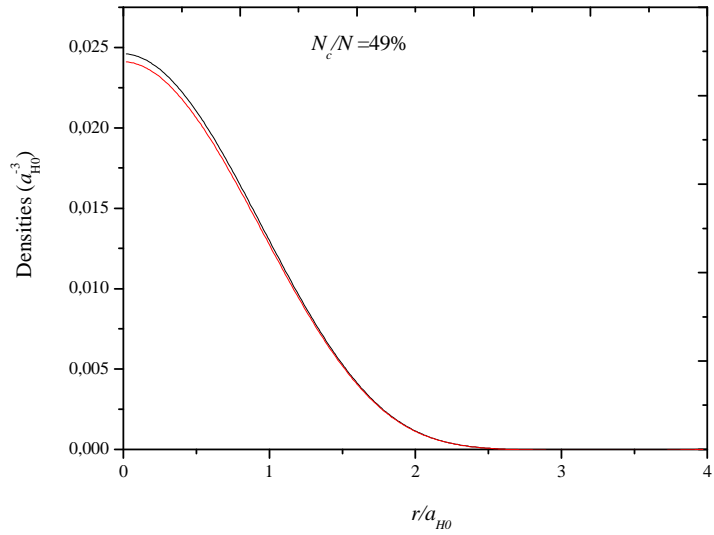

**Figure 7. Same as figure 6 for $N_c/N = 49\%$.**

**Variational self-consistent theory for trapped Bose gases at finite temperature**

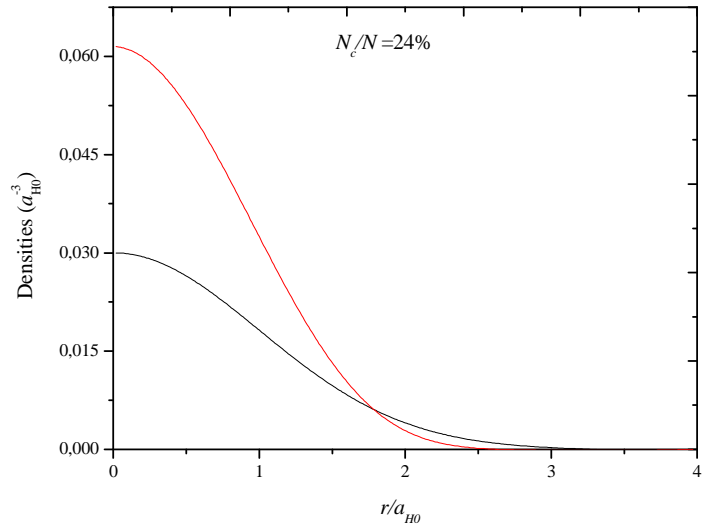

**Figure 8.** Same as figure 6 for $N_c/N = 24\%$.

## 6. Conclusions and Perspectives

By using a gaussian density operator, we derive from the time-dependent variational principle of Balian-Vénéroni a set of coupled equations of motion for a self-interacting trapped bose gas. These equations govern in a self-consistent way the dynamics of the condensate, the thermal cloud and the anomalous average. Our time dependent Hartree-Fock-Bogoliubov (TDHFB) equations generalize in a natural way many of the famous approximations found in the literature such as the Bogoliubov [3], the Gross-Pitaevskii[5], the Popov[31], the Beliaev [32] and the Bogoliubov-de Gennes equations[11-12].

In order to apprehend better the advantages of our approach, we solve numerically the static TDHFB equations in the local limit for a contact potential and a harmonic trap. The outcomes of our numerical explorations are numerous. First of all, the numerical resolution of our equations is relatively easy and is not as time-consuming as the HFB-BdG calculations especially for large atom numbers. For instance, the latter cannot be handled correctly as soon as $N \sim 10^4 - 10^5$. By contrast, there are no such limitations in our case. Secondly, although we obtain for the condensate density a quite good agreement with the literature and with the experiments, our predicted thermal cloud and anomalous average differ substantially from the HFB-BdG results

# Variational self-consistent theory for trapped Bose gases at finite temperature

especially near the center of the trap. Indeed, we obtain gaussian shapes for both densities while the HFB-BdG calculations predict ''holes'' (that is local minima) near the center. These holes are absent (at least concerning $\tilde{n}$) in the experiments of Gerbier and al.[29] and Caracanhas and al.[23]. We recall that the HFB-BdG numerical calculations have been performed for a reduced atom number ($N \sim 2000$)[12]. The question that naturally arises is whether these holes may disappear for large atom numbers. In fact, a preliminary HFB-BdG calculation seems to confirm this conclusion. We plan to publish more details in the near future [33].

On the other hand, owing to its importance to account for many-body effects, we have analyzed the behavior of the anomalous density. We recover a well-known theoretical prediction of HFB-BdG[12] since $\tilde{m}$ increases with the temperature and then decreases as one approaches the transition. Moreover, we show that at low temperatures, the anomalous density is greater than the non condensate density, although both are very small. The former necessarily plays a major role in the condensation phenomenon. Any approach neglecting the anomalous average at low temperatures will inevitably lead to inconsistencies [30].

Finally, as for any variational approach, one has to discuss the relevance of the trial space for the problem at hand and how to extend it in order to take into account many important effects which we did not deal with in this paper. Among the most important phenomena, we can cite in particular the effect of the inclusion of the triplets, the damping of the excitations (Landau and Beliaev) and the particle exchange. While the former may be easily handled by the ''post-gaussian'' approximation, the two latter require dissipative effects which are totally absent in the present formalism (recall that the von-Neumann entropy is conserved).

**Acknowledgments**

We would like to thank W. Ketterle, G.V. Shlyapnikov, D. Mathew, V.I. Yukalov and U. Al Kawaja for fruitful discussions.

# Variational self-consistent theory for trapped Bose gases at finite temperature